\documentclass[pra,showpacs]{revtex4}
\usepackage[applemac]{inputenc}
\usepackage{amssymb,amsmath}
\usepackage[mathscr]{euscript}
\usepackage{graphicx}

\newcommand{\kB}{k_{\mathrm{B}}}
\newcommand{\ave}[1]{\left \langle {#1} \right \rangle}
\newcommand{\myl}{\mathscr{L}}
\newcommand{\myh}{\mathscr{H}}

\begin{document}

\preprint{arXiv:quant-ph/0211072}

\title{A simple quantum heat engine}

\author{Jacques Arnaud}

\email{arnaudj2@wanadoo.fr}

\affiliation{Mas Liron, F30440 Saint Martial, France}

\author{Laurent Chusseau}

\email{chusseau@univ-montp2.fr}

\homepage{http://www.opto.univ-montp2.fr/~chusseau}

\affiliation{Centre d'Électronique et de Micro-optoélectronique de Montpellier, UMR 5507 CNRS, Université Montpellier
II, F34095 Montpellier, France}

\author{Fabrice Philippe}

\altaffiliation[Also at ]{MIAp, Université Paul Valéry, F34199 
Montpellier, France}

\email{Fabrice.Philippe@univ-montp3.fr}

\affiliation{Laboratoire d'Informatique de Robotique et de Microélectronique de Montpellier, UMR 5506 CNRS, 161 Rue Ada,
F34392 Montpellier, France}

\date{\today}

\begin{abstract}      
Quantum heat engines employ as working agents multi-level systems instead of gas-filled cylinders. We consider
particularly two-level agents such as electrons immersed in a magnetic field. Work is produced in that case when the
electrons are being carried from a high-magnetic-field region into a low-magnetic-field region. In watermills, work is
produced instead when some amount of fluid drops from a high-altitude reservoir to a low-altitude reservoir. We show
that this purely mechanical engine may in fact be considered as a two-level quantum heat engine, provided the fluid is
viewed as consisting of $n$ molecules of weight one and $N-n$ molecules of weight zero. Weight-one molecules are
analogous to electrons in their higher energy state, while weight-zero molecules are analogous to electrons in their
lower energy state. More generally, fluids consist of non-interacting molecules of various weights. It is shown that,
not only the average value of the work produced per cycle, but also its fluctuations, are the same for mechanical
engines and quantum (Otto) heat engines. The reversible Carnot cycles are approached through the consideration of
multiple sub-reservoirs. 
\end{abstract}


\maketitle

\section{Introduction}

Quantum heat engines employ as working agents multi-level systems instead of gas-filled cylinders. We consider
particularly two-level agents such as electrons immersed in a magnetic field, work being produced when the electrons are
being carried from high-magnetic-field regions into low-magnetic-field regions. The purpose of the present paper is to
show that quantum heat engines may be viewed as purely mechanical systems such as watermills. This is achieved by
considering the microscopic content of the fluid reservoirs. To wit, these reservoirs are viewed as collections of
molecules of various weights. In the simplest configuration, the molecule weights are supposed to be equal to zero or
one, with the average molecular weight corresponding to the fluid density. For such models, watermills correspond to
quantum (Otto) heat engines employing as working agent two-level systems. Not only the average work produced per cycle,
but also the work fluctuations, are the same.

The operation of quantum heat engines that employ as working agents multi-level systems, for example harmonic
oscillators, free particles in a box, three-level atoms or electrons submitted to magnetic fields, were introduced by
Geva and Kosloff in \cite{geva,feldmann}, see also \cite{he,scully}.  The present paper, restricted to slow elementary
systems, should help students get a feel for the method of extraction of mechanical energy from heat baths without
having to enter into Thermodynamics concepts in a first approach. 

In the following we consider in turn a classical mechanical system (watermill), a classical Otto heat engine and a
quantum Otto heat engine. We explain in what sense the first is related to the third. Through the introduction of
multiple reservoirs, the simple Otto cycle leads to the reversible Carnot cycle, which, inherently, exhibits greater
efficiency.

\begin{enumerate}

\item Let us begin by considering a purely mechanical system, namely a watermill. As is well known, the energy released
by water dropping from a high-level reservoir into a lower one may be employed to lift a weight. Ideally (e.g.,
neglecting friction) the efficiency is unity in the sense that the potential energy lost by water is entirely converted
into the weight potential energy (because the motion is assumed to be slow no consideration is given here to kinetic
energy). However, we may also define the efficiency $\eta$ as the ratio of the work actually performed to the one that
would have been performed if the same amount of water had been dropped to some lower reference level. If the reservoir
altitudes with respect to that reference level (see Fig.~\ref{figroue}) are denoted by $\epsilon_{l}>0$ and
$\epsilon_{h}>\epsilon_{l}$, where the subscripts $l$ and $h$ stand for ``low'' and ``high'', respectively, we have
obviously for the ideal engine $\eta=1-\epsilon_{l}/\epsilon_{h}$. We will show that this expression for the efficiency
applies as well to quantum Otto heat engines. \linebreak
To facilitate the comparison, we in fact consider a slightly different kind of mill. The high-altitude reservoir is
supposed to contain a high-density fluid and the low-altitude reservoir a low-density fluid. Work is produced when some
volume of fluid is exchanged between the two reservoirs. Let the fluids be modeled as collections of $N$ balls (or
molecules), $n$ of them having weight 1 and $N-n$ having weight 0. We suppose for simplicity that the total number of
balls, $N$, is the same for the two reservoirs. The number of weight-1 balls in the lower reservoir is denoted by
$n_{l}$ and the number of weight-1 balls in the higher reservoir by $n_{h}$. Clearly, the exchange of two
randomly-selected ball between the reservoirs will produce a positive average work if $n_{h}>n_{l}$. As far as the
average work is concerned only fluid densities matter. Other microscopic models with $n_{0}$ balls of weight $w_{0}$,
$n_{1}$ balls of weight $w_{1}$, and so on, would do just as well, as long as the \emph{density} $w/N$ remains the same.
Here, $w\equiv \sum n_{k}w_{k}, k=0,1,2...$ denotes the total fluid weight. The microscopic model determines, however,
the \emph{fluctuations} of the work produced. In this paper, we evaluate both the average work produced and the variance
of that work, averages being defined with respect to an ensemble of initially identical engines. Only the most
elementary rule of probability is needed, namely, the fact that when an urn contains $N$ balls possessing labels that
help distinguish one ball from another, the probability of picking up any particular ball is $1/N$.

\begin{figure}
\begin{center}
\includegraphics[scale=0.7]{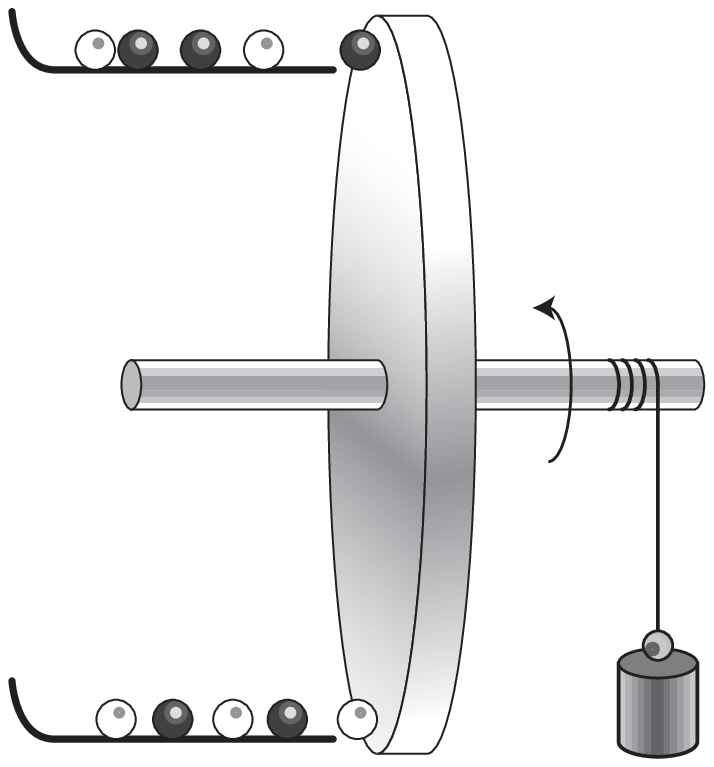}
\end{center}
\caption{Schematic representation of a \emph{classical} mechanical engine that converts the potential energy of heavy
balls into the lift of a weight. In the figure the two reservoirs contain $N=5$ balls each. The lower one is 
located at altitude $\epsilon_{l}$. The upper one is located at altitude $\epsilon_{h}$. A cycle consists of exchanging
two balls with the help of a rotating plate. One may suppose that the balls are made up of steel and that they stick to
a magnetic circular plate.}
\label{figroue}
\end{figure}

\item Let us explain next what are, in essence, \emph{classical} Otto heat engines. The working agent may be a
gas-filled cylinder with a movable piston, initially in thermal contact with a hot bath. When the cylinder is separated
from the bath and allowed to expand in length, the piston motion may be employed to lift a weight, while the gas cools
down. As a second step, the cylinder is put in contact with a cold bath. The cylinder is subsequently separated from the
bath and allowed to shrink back to its original length. Note that, in contrast with the four-stroke Carnot cycle, the
cylinder length does not vary when it is in contact with the baths. Heat-engine efficiencies, $\eta$, are defined in
general as ratios of the work performed and the heat removed from the hot bath. The Otto cycle efficiency may reach the
Carnot efficiency $\eta_{C}=1-T_{cold}/T_{hot}$, where $T_{cold,hot}$ denote the bath temperatures, only in the limit
where the work produced is vanishingly small.

\item In \emph{quantum} Otto engines the working agent, instead of being a classical gas, is a multilevel system, for
example (spin-1/2) electrons immersed in a magnetic field $B$, for the two-levels case. 
Electrons possess different energies depending on whether their magnetic moments point in the magnetic field direction
or in the opposite direction. For simplicity, we select units such that the level energy difference $\epsilon$ is equal
(rather than just proportional) to $B$. Without loss of generality, the electron lower-level energies are set equal to 0
\cite{note1}. The upper-level energy is then simply $\epsilon=B$. If an electron in the upper state is transferred from
a high-magnetic-field region into a low-magnetic-field region the electron energy gets reduced. This reduced energy is
delivered to the agent that helps displace the electron. For example, if the electron is acted upon by an electric field
the energy is delivered to the battery that generates that field. \linebreak
In the quantum Otto cycle an electron (the working agent), initially in thermal contact with a bath at temperature
$T_{h}$ and immersed in a magnetic field $B_{h}=\epsilon_{h}$, gets carried to a bath at a lower temperature $T_{l}$ and
lower magnetic field $B_{l}=\epsilon_{l}$. This electron is subsequently carried back to the original bath, thereby
closing the cycle, with some net work being performed. Because  the cycle is assumed to be arbitrarily slow, there is
enough time for the electron to reach a regime of thermal equilibrium with the baths. The two processes ($h\to l$ and
$l\to h$) being statistically independent it is easy to evaluate the statistics of the net work performed, and in
particular the average work and the work variance.

\end{enumerate}

\begin{figure}
\begin{center}
\includegraphics[scale=0.5]{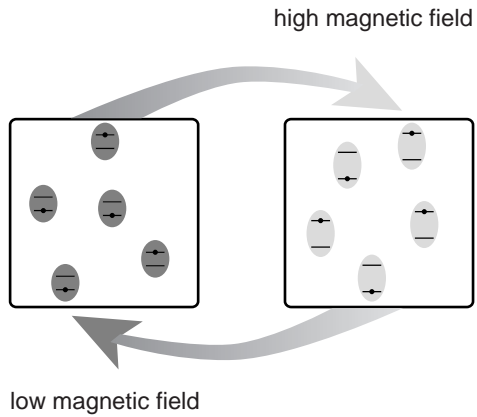}
\end{center}
\caption{Schematic representation of a \emph{quantum} (spin-1/2) Otto heat engine. The arrows indicate that electrons
are exchanged between the high magnetic field, high temperature, bath and the low magnetic field, low temperature,
bath.}
\label{figmagnet}
\end{figure}

To facilitate the comparison that we wish to make, two slight modifications of the quantum Otto cycle just described are
made. First, instead of a single electron moving back and forth between the two baths we suppose that, at the same time,
an electron in equilibrium with the hot bath is carried to the cold bath and an electron in equilibrium with the cold
bath is carried to the hot bath, as is illustrated in Fig.~\ref{figmagnet}. The two arrows illustrate the exchange of two electrons.

In the above discussion the bath microscopic nature has been left unspecified. We now assume that the heat baths are
actually collections of two-level electrons, similar to the ones previously referred to as being the ``working agents''.
These electrons are supposed to interact weakly with one another so that an equilibrium situation may eventually be
reached, but the interaction energy is so small that the total energy is very nearly the sum of individual electron
energies. Let an (isolated) bath contain $n$ electrons in the upper state and $N-n$ in the lower state. If an electron
is being picked up from the bath at random, the probability that it be in the upper state is \emph{exactly} $n/N$. At
that point we notice that there is an exact correspondence with the fluid previously considered, provided  the latter is
modeled as a collection of distinguishable balls, $n$ of them having weight 1, and $N-n$ of them having weight 0. 
In the limit that $n$ and $N-n$ are both large a temperature may be defined. The quantity $n/N$ previously called
``density'' is equal to $1/[\exp(\epsilon/ \kB T)+1]$, where $\kB$ denotes the Boltzmann constant, see
Appendix~\ref{reservoir}. Notice that this temperature is negative if the ``density'' exceeds one-half \cite{note2}. The
point we wish to make here is that the comparison made between particular mechanical and  heat engines is \emph{exact},
while the concept of temperature may be introduced only in the limit of large particle numbers.
 
The average and variance of the work produced by modified watermills are evaluated in Section~\ref{mechanical}. A
generalization is made to the case where the high and low reservoirs are split into a number of sub-reservoirs for the
purpose of comparing mechanical systems and Carnot heat engines. The treatment of quantum heat engines is given in
Section~\ref{quantum} for single low- and high-temperatures baths and in Section~\ref{multiple} for multiple cold and
hot baths.  

Only two-level systems are considered in detail. But note that when the ball weights are allowed to be $w_{k}=k$, where
$k=0,1,2...$, the mechanical system is equivalent to quantum heat engines employing as working agents one-dimensional
oscillators, whose level energies are, as is well-known, $\epsilon_{k}=(k+1/2)\hbar \omega$, where $\hbar$ denotes the
Planck constant divided by $2\pi$ and $\omega$ denotes the oscillator angular frequency. The latter plays for
oscillators a role similar to the magnetic field $B$ for electrons. Previous results \cite{arnaud} are recovered in that
manner. 

From the quantum point of view, ideal gases confined to one-dimensional boxes of length $L$ are collections of
non-relativistic particles of masses $m$, whose wavefunctions must vanish at $x=0$ and $x=L$. The energy levels are in
that case given by $\epsilon_{\ell}=p^2/2m$, where $p=\hbar k$, $kL=\ell \pi$, $\ell =1,2...$. That is, the energy
levels are of the form $\epsilon_{\ell}=\ell^2/L^2$, the quantity $1/L^2$ playing therefore for non-relativistic
particles a role similar to the magnetic field for electrons. In our comparison, ideal gases would correspond to fluids
whose molecular weights would be 1,4,9.... Ideal gases will not be considered further in the present paper.

\section {The mechanical engine}
\label{mechanical}

Consider the mechanical engine shown in Fig.~\ref{figroue} that converts the energy released by dropping heavy balls
from one level to a lower one into the energy required to lift a weight. Only slow regular (quasi-static) motions are
contemplated so that kinetic energies may be ignored. Specifically, the engine consists of two reservoirs, a lower one
at altitude $\epsilon_{l}>0$ defined with respect to some reference level, an a higher one at altitude 
$\epsilon_{h}>\epsilon_{l}$. The two reservoirs are supposed to contain $N$ balls each, of various weights. The total
weight of the lower-reservoir balls is denoted by $w_{l}$ and the total weight of the higher-reservoir balls by $w_{h}$.

Let now two randomly-selected balls be exchanged between the reservoirs. As is illustrated in the figure this could be
accomplished by a rotating plate, but the details of the mechanism need not be given as we are only concerned in
establishing matters of principle. Since the probability of picking up a particular ball from a reservoir is $1/N$, the
average energy removed from the lower reservoir is $\epsilon_{l}w_{l}/N$ and the average energy added to the higher
reservoir is $\epsilon_{h}w_{l}/N$. At the same time, a ball is being picked up at random from the higher reservoir and
carried to the lower reservoir so that the numbers of balls in each reservoir remain equal to $N$. An average energy
$\epsilon_{h}w_{h}/N$ is thus removed from the higher reservoir and an average energy $\epsilon_{l}w_{h}/N$ added to the
lower one. 

According to the above discussion the average energies in the lower and higher reservoirs are incremented respectively
by
\begin{subequations}
\label{Q}
\begin{align}
\ave{Q_{l}} &= \epsilon_{l}\frac{w_{h}}{N}-\epsilon_{l}\frac{w_{l}}{N}=\epsilon_{l}\frac{w_{h}-w_{l}}{N},\\
\ave{Q_{h}} &= \epsilon_{h}\frac{w_{l}}{N}-\epsilon_{h}\frac{w_{h}}{N}=-\epsilon_{h}\frac{w_{h}-w_{l}}{N}.
\end{align}
\end{subequations}
The average work performed follows from the law of conservation of energy
\begin{equation}
\label{W}
\ave{W} = -\ave{Q_{l}}-\ave{Q_{h}}=(\epsilon_{h}-\epsilon_{l})\frac{w_{h}-w_{l}}{N}. 
\end{equation}
Since by convention $\epsilon_{h}>\epsilon_{l}$ we require that $w_{h}>w_{l}$ if the system is aimed at \emph{lifting}
the weight, and $w_{h}<w_{l}$ in the opposite situation.  

The engine efficiency $\eta$ is defined as the ratio of the average work performed $ \ave{W}$ (see above) and the
average higher-reservoir energy consumption $-\ave{Q_{h}}$. We obtain
\begin{equation}
\label{eta}
\eta = \frac{ \ave{W}}{- \ave{Q_{h}}} = 1+\frac{\ave{Q_{l}}}{\ave{Q_{h}}} = 1-\frac{\epsilon_{l}}{\epsilon_{h}}.
\end{equation}

\begin{figure}
\begin{center}
\includegraphics[scale=1]{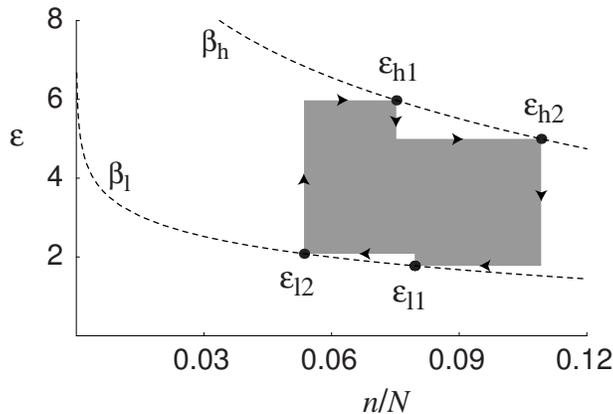}
\end{center}
\caption{Schematic representation of a heat engine. The cold and hot reservoirs have been split into two sub-reservoirs
each ($m=2$) in order to improve the engine efficiency. The Carnot efficiency would be reached in the limit of a large
number of sub-reservoirs ($m \to \infty$). The reservoirs are shown as dots with $n/N$ in abscissa, where $n$ represents
the number of balls with weight one and the altitude $\epsilon$ in ordinate. The work performed is equal to the cycle
area. Note that in comparison with the usual representation of Thermodynamics cycles $\epsilon$ corresponds to volume
and $n$ to pressure. The volume is represented here on the vertical rather than horizontal axis as usual because of the
physical interpretation attached to $\epsilon$ in our model.}
\label{figcycle}
\end{figure}

Let us now suppose that there are $m$ low-level reservoirs (referred to as sub-reservoirs) at altitudes $\epsilon_{l,1},
\epsilon_{l,2}, ... \epsilon_{l,m}$, with weights $w_{l,1}, w_{l,2}, ... w_{l,m}$, respectively. (Comas separating $l$
from the numbers 1,2... that label the sub-reservoirs will henceforth be suppressed for brevity, except when a confusion
may arise). Likewise, we suppose that there are $m$ high-level sub-reservoirs at altitudes $\epsilon_{h1},
\epsilon_{h2}, ... \epsilon_{hm}$, with weights $w_{h1}, w_{h2}, ... w_{hm}$. Randomly selected balls are carried at the
same time from the sub-reservoir $l1$ to the sub-reservoir $l2$,\ldots, from $lm$ to $h1$, $h1$ to $h2$\ldots, and $hm$
to $l1$, thereby closing the cycle, as illustrated in Fig.~\ref{figcycle} for the case where $m=2$. Using the same
argument as above, the total average energy increment in the lower and higher reservoirs is
\begin{subequations}
\label{betaq}
\begin{align}
N\ave{Q_{l}} &= \epsilon _{l1}(w_{hm}-w_{l1})+\ldots+\epsilon_{lm}(w_{l,m-1}-w_{lm})\nonumber \\
 & =\epsilon_{l1}w_{hm}+\sum_{i=1}^{m-1}w_{li}(\epsilon_{l,i+1}-\epsilon_{li})-\epsilon_{lm}w_{lm} ,\\
N\ave{Q_{h}} &=\epsilon _{h1}(w_{lm}-w_{h1})+\ldots+\epsilon_{hm}(w_{h,m-1}-w_{hm})\nonumber \\
 & =\epsilon_{h1}w_{lm}+\sum_{i=1}^{m-1}w_{hi}(\epsilon_{h,i+1}-\epsilon_{hi})-\epsilon_{hm}w_{hm} .
\end{align}
\end{subequations}
The average work produced is $\ave{W}=-\ave{Q_{l}}-\ave{Q_{h}}$. In the next Section we consider the limit in which $m$
goes to infinity while the steps in $\epsilon$ tend to zero as one goes from one sub-reservoir to the next. 

As far as the average work performed is concerned, only the total ball weights in the sub-reservoirs are relevant.
However, if we are interested in the \emph{statistics} of the work performed, we need specify the individual balls
weights. For simplicity, let us suppose that the sub-reservoir $k$ contains $n_{k}$ balls of weight 1 and $N-n_{k}$
weightless balls, with $k$ running from 1 to $2m$, with the understanding that $2m+1\equiv 1$. The probability that a
ball of weight 1 be picked up from the sub-reservoir $k$ is $f_{k}=n_{k}/N$. If that event occurs, the work performed
$W$ is incremented by $ \epsilon_{k}-\epsilon_{k+1}$ according to the law of conservation of energy, since the ball
picked up from the sub-reservoir $k$ is being transferred to the sub-reservoir $k+1$. If the event does \emph{not} occur
$W$ is unchanged. Since the ball selections from different sub-reservoirs are independent the contributions of the
individual events averages and variances add up, and we obtain
\begin{equation}
\label{variance}
\frac{\mathrm{variance}(W)}{\ave{W}}=\frac{\sum (\epsilon_{k}-\epsilon_{k+1})^2 f_{k}(1-f_{k})}
{ \sum (\epsilon_{k}-\epsilon_{k+1}) f_{k}}, 
\end{equation}
where the sums run from $k=1$ to $2m$. It follows from this expression that if the altitude difference between the 
highest and lowest sub-reservoirs is kept constant, and in the limit that $m\to \infty $, the work performed is essentially
non-fluctuating. Indeed, if $\epsilon_{k}-\epsilon_{k+1}$ is of order $\Delta$ the numerator, of order $m\Delta^2$,
tends to zero when the denominator, of order $m\Delta$, is kept constant. A similar situation occurs for some quiet-pump
lasers\cite{arnaud2}.

\section{The equivalent quantum heat engine}
\label{quantum}

The mechanical engine described in the previous Section may be viewed as a quantum heat engine, a ball of weight $w$
located in a reservoir at altitude $\epsilon$ being viewed as the occupation of the level $\epsilon w$ of a multilevel
system.

Suppose for example that the ball weights may only be 0 or 1, and consider a 2-level atom whose ground state energy is
set equal to zero and whose excited state energy is $\epsilon$ through appropriate normalization. Weight-0 balls are
equivalent to such atoms in the ground state while weight-1 balls are equivalent to such atoms in the excited state.
Accordingly, each reservoir may now be viewed as a collection of $N$ two-level atoms, $n$ of them being in the excited
state and $N-n$ of them in the ground state. Changing the reservoir altitude amounts to changing the atom level-energy
spacing. As an example, let us suppose that the 2-level system is actually an electron immersed in a magnetic field. A
change of the reservoir altitude corresponds in that case to a change in magnetic field strength. The two heat baths are
represented in Fig.~\ref{figmagnet}. To avoid a confusion, let us emphasize that the reservoirs are \emph{not} viewed
as reservoirs of structureless molecules.

We thus consider a collection of $N$ atoms with level energies 0 and $\epsilon$. The inverse temperature $\beta\equiv
1/ \kB T$ of such an atomic collection may be found in every text-book of statistical mechanics. For the reader
convenience the derivation is recalled in Appendix~\ref{reservoir}. If the total energy is $n\epsilon$ 
\begin{equation}
\label{beta}
\beta=\frac{\ln(\frac{N}{n}-1)}{\epsilon} .
\end{equation}

Using as before subscripts $l$ for the lower reservoir and subscripts $h$ for the higher one, we have therefore
\begin{subequations}
\label{betalh}
\begin{align}
\beta_{l} =\frac{\myl}{\epsilon_{l}} , &\quad \myl \equiv \ln(\frac{N}{n_{l}}-1) ,\\
\beta_{h} =\frac{\myh}{\epsilon_{h}} , &\quad \myh \equiv \ln(\frac{N}{n_{h}}-1) ,
\end{align}
\end{subequations}
respectively.  Note that $\beta_{l,h}$ are positive when $n_{l,h}$ are smaller than $N/2$ and negative when they are
larger than $N/2$. From the previous expressions, the  average work produced, given in \eqref {W}, may be written 
as 
\begin{equation}
\label{work}
W(\epsilon_{l},\epsilon_{h})= \left( \epsilon_{h} -\epsilon_{l} \right) \left(
\frac{1}{\exp(\beta_{h}\epsilon_{h})+1}-\frac{1}{\exp(\beta_{l}\epsilon_{l})+1} \right) ,
\end{equation}
where we consider that the reservoir temperatures have been ascribed some values.

\begin{figure}
\begin{center}
\includegraphics[scale=1]{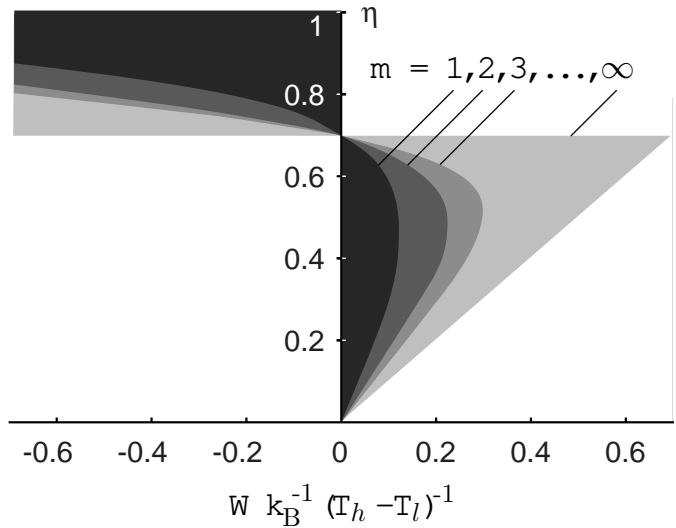}
\end{center}
\caption{This figure shows how the maximum heat-engine (right-hand-side) heat-pump (left-hand-side) efficiencies evolve
as a function of the average work performed per cycle $W$ in the case of $m$=1 (Otto cycle), $m=2$, $3$, and an infinity
(Carnot cycle) of sub-reservoirs. We have set $\beta_{c}=1.38$, $\beta_{h}=0.42$. The grey area ($m=1$) corresponds to
$\epsilon$-values varying from 0 to $\infty$. The coefficient of performance (COP) of heat pumps is equal to the inverse
of the efficiency.}
\label{figeta}
\end{figure}

As an example of a heat engine suppose that $\epsilon_{l}=1$ and $\epsilon_{h}=2$. The calculated efficiency is
$\eta=0.5$. Further, for $N=10\,000$, $n_{l}=2\,000$ and $n_{h}=3\,000$, the work delivered per cycle is $W=0.1$ in the
energy unit employed to define $\epsilon$. For the above numerical values, we calculate that $\beta_{l}=1.38$, 
$\beta_{h}=0.42$. The maximum (Carnot) efficiency would be $\eta_{C}\equiv 1-\beta_{h}/\beta_{c}=0.69$. If we keep
constant $\beta_{l}$ and $\beta_{h}$ and vary $\epsilon_{l}$, $\epsilon_{h}$, the work performed given in 
\eqref{work} may reach 0.2. Figure~\ref{figeta} (curve $m=1$) gives the maximum efficiency $\eta$ for a given value of the average
work $\ave{W}$ performed in a cycle. Positive work corresponds to a heat engine while negative work corresponds to a
heat pump, whose coefficient of performance is the reciprocal of $\eta$. The gray areas have been obtained by randomly
selecting positive $\epsilon$-values. 

In subsequent examples we keep $N=10\,000$ and the absolute value of the population difference: $\left\lvert n_{l}-n_{h}
\right\rvert =1\,000 $.  To exemplify heat pumps, it suffices to exchange the values of $n_{l}$ and $n_{h}$, keeping
$\epsilon_{l}$ and $\epsilon_{h}$ the same as above.  The efficiency, as defined earlier, remains $\eta=0.5$. 
However, it is usual (and natural) in the case of heat pumps to define a coefficient of performance (COP) as the ratio
of the amount of heat generated in the hot reservoir (e.g., in the house) to the work required to operate the system
(e.g., from the power line).  The COP (reciprocal of $\eta$) is in our example equal to 2.  We calculate further that
$\beta_{l}=0.85$, $\beta_{h}=0.69$ and thus the maximum (Carnot) heat-pump COP is $1/\eta_{C}=5.5$.

Negative reservoir temperatures are now particularly considered. If the number of balls of weight 1 in the reservoir at
altitude $\epsilon=1$ is equal to 7\,000 and the number of balls of weight 1 in the reservoir at altitude $\epsilon=2$ is
equal to 8\,000 the reservoir temperatures are both negative. The $\beta$, $Q$ and $W$-values are simply opposite in sign
to the previous ones, with $\beta_{l}=-0.69$, $\beta_{h}=-0.85$. It follows that the reservoir at altitude $\epsilon=2$
is colder than the reservoir at altitude $\epsilon=1$.  The system is a heat pump with some work flowing into it and
heat being transferred to the hotter reservoir. If the above values of $n_{l}$ and $n_{h}$ are interchanged the system
becomes a heat engine indistinguishable from the one initially considered as far as the operating parameters are
concerned. If $n_{l}=4\,500$, $n_{h}=5\,500$ and $\epsilon_{c}=1$, $\epsilon_{h}=2$ the cold reservoir temperature is
positive while the hot reservoir temperature is negative, with $\beta_{l}=0.2$, $\beta_{h}=-0.1$.  As before, the heat
engine delivers an average energy $\ave{W}=0.1$ and the efficiency is $\eta=0.5$. The maximum attainable efficiency is
unity.

For the sake of comparison with the case of electrons submitted to a magnetic field, note that the quantity denoted here
$n/N$ corresponds to the electron spin (divided by $\hbar$) plus 1/2.  The reservoir altitude $\epsilon$ corresponds to
the quantity denoted by $\omega$ in \cite{geva}. The relation between $n/N$ and the inverse temperature $\beta$ given in
\eqref{beta} coincides with the one given in that reference. 

\section{Multiple sub-reservoirs}
\label{multiple}

The engine efficiency may improve if a large number, $m$, of low and high sub-reservoirs are employed, as shown in
Fig.~\ref{figcycle} for the case where $m=2$. The number of balls in each sub-reservoir is again equal to $N$.  The low
sub-reservoirs are labeled by $lk$, while the high sub-reservoirs are labeled by $hk$ where $k=1,\ldots,m$. The total
weight of the balls in sub-reservoir $lk$ with altitude of $\epsilon_{lk}$ is denoted by $n_{lk}$ and the total weight
of the balls in sub-reservoir $hk$ with altitude of $\epsilon_{hk}$ is denoted by $n_{hk}$.

We now suppose that $n_{lk}\equiv n_{l}(\epsilon_{lk})$ and $n_{hk}\equiv n_{h}(\epsilon_{hk})$, where we have
introduced the functions
\begin{equation}
\label{c}
   n_{l}(\epsilon) =N f(\beta_{l}\epsilon) , \quad
   n_{h}(\epsilon) =N f(\beta_{h}\epsilon), 
\end{equation}
where $\beta_{l,h}$ are constants and $f(x)$ denotes a function of $x$ to be later specified. 

The above expressions of $\ave{Q_{l}},~\ave{Q_{h}}$ are now written as
\begin{subequations}
\label{newbetaq}
\begin{align}
\beta_{l}\ave{Q_{l}} &=c_{1}f \left( \myh_{m} \right) +\sum_{i=1}^{m-1}f \left( \myl_{i} \right)  \left( \myl_{i+1}-\myl_{i}
\right) -\myl_{m}f \left( \myl_{m} \right)  , \\
\beta_{h}\ave{Q_{h}} &=h_{1}f \left( \myl_{m} \right) +\sum_{i=1}^{m-1}f \left( \myh_{i} \right)  \left( \myh_{i+1}-\myh_{i}
\right) -\myh_{m}f \left( \myh_{m} \right)  ,
\end{align}
\end{subequations}
where we have defined $\myl_{k}\equiv \beta_{l}\epsilon_{lk}$ and $\myh_{k}\equiv \beta_{h}\epsilon_{hk}$. They enable us
to evaluate the average work performed $\ave{W}=-\ave{Q_{l}}-\ave{Q_{h}}$ and the efficiency
$\eta=-\ave{W}/\ave{Q_{h}}$ for given values of the parameters $\myl_{k}$, $\myh_{k}$, $k=1,2\ldots m$. 

If the $\epsilon$-values do not vary much from one $k$-value to the next, the sums may be converted into integrals, and
we obtain \cite{arnaud}
\begin{subequations}
\label{newbetaq'}
\begin{align}
\beta_{l}\ave{Q_{l}}=s \left( \myl_{1},\myh_{m} \right) -s \left( \myl_{m} \right)  , \\
\beta_{h}\ave{Q_{h}}=s \left( \myh_{1},\myl_{m} \right) -s \left( \myh_{m} \right)  ,
\end{align}
\end{subequations}
where
\begin{equation}
\label{s}
s(x,y) \equiv xf(y)-\phi(x) , \quad \phi (x) \equiv \int_{}^{x}f(x') dx' , \quad s(x)\equiv s(x,x).
\end{equation}
The lower limit of the integral is unimportant.
 
The simpler expressions given in Section~\ref{mechanical} corresponding to $m=1$ are equivalent to the so-called Otto
cycle of heat engines. It is recovered from the present formulation by setting $\myl_{1}=\myl_{m}$ and $\myh_{1}=\myh_{m}$.
If, on
the other hand, $\myh_{m}=\myl_{1}$ and $\myl_{m}=\myh_{1}$ the quantity  $\beta_{l}Q_{l}+\beta_{h}Q_{h}$ vanishes, and the
Carnot efficiency is reached. The average work performed may be written in that case as
\begin{equation}
\label{work'}
\ave{W}= \kB \left( T_{h}-T_{l} \right)  \left( s \left( \myl_{1} \right) -s \left( \myl_{m} \right)  \right) , 
\end{equation}
where we have introduced the absolute reservoir temperatures $T_{l}$, $T_{h}$. The average work is the product of the change of
temperature and the change of entropy. 
 
To proceed further, we must specialize the above results. Consider the case where the sub-reservoirs contain $n$ balls
of weight unity, the other balls being weightless as discussed in Section~\ref{mechanical}. According to
Appendix~\ref{reservoir} the appropriate function is in that case
\begin{equation}
\label{ch}
f(x)= \frac{1}{\exp(x)+1} \Longrightarrow s(x,y)= \frac{x}{\exp(y)+1}+\ln(1+\exp(-x)). 
\end{equation}
Using the above expression of the entropy, we find that $W_{max}=\kB (T_{h}-T_{l})\ln(2)$. More generally, when the
allowed weight-values are $0$, $1\ldots$, $2j$ we have $W_{max}=\kB \left( T_{h} - T_{l} \right) \ln(2j+1)$
\cite{scully}.
When $\beta_{l}=1.38$, $\beta_{h}=0.42$ ($\kB T_{h}- \kB T_{l}=1.66$) the cycle is reversible and the Carnot efficiency
$\eta_{C}=0.695$ is reached when for example $\epsilon_{l1}=1$, $\epsilon_{lm}=1.1$, $\epsilon_{h1}=3.6$,
$\epsilon_{hm}=3.3$. In that case $W=0.05$. When we allow the $\epsilon$ to vary the work may reach the value
$W_{max}=1.14$. Figure~\ref{figeta} shows how the heat-engine maximum efficiency varies as a function of the work
performed for 1 (Otto cycle), 2, 3, and an infinity (Carnot cycle) of cold and hot sub-reservoirs. Positive work
corresponds to heat engines and negative work to heat pumps, whose coefficients of performance are the reciprocals of
$\eta$. The figure gives also the \emph{minimum} value of the efficiency. Curiously, the \emph{minimum} Carnot cycles
efficiency is proportional to $W$.

The above formulas are valid also when the hot reservoir has a negative temperature, in which case the maximum
efficiency is unity. For example, if $\beta_{l}=0.2$,  $\beta_{h}=-0.1$, we obtain for
$\epsilon_{l1}=\epsilon_{lm}=\epsilon_{hm}=0.3$, $\epsilon_{h1}=14$, a work produced $W=2.5$ and an efficiency
$\eta=0.997$.

\section {Conclusion}
   
We have seen that a simple urn model suffices to obtain expressions for the average value and the variance of the work
produced by a simple mechanical system consisting of balls dropping from one level to another. We have shown that this
system is formally the same as a quantum heat engine employing as a working substance multi-level atoms. When
temperatures are ascribed to the two reservoirs according to the laws of Quantum Statistical Mechanics, the expression
for the Carnot efficiency is obtained in the limit of a large number of sub-reservoirs. In particular, when the ball
weights may only be zero or one, the mechanical system is equivalent to a quantum heat engine employing as working
substance electrons submitted to a spatially varying magnetic field. When the ball weights may take any non-negative
integral values, the system describes heat engines employing linear oscillators as working substance.    

\appendix
\section{Reservoir temperature}
\label{reservoir}

We have considered in the main text a reservoir at altitude $\epsilon$ containing $N$ distinguishable balls of weight
$0$ or $1$. These balls were likened to two-level atoms in the ground state and the excited state, respectively. The
purpose of the present appendix is to recall how temperatures may be assigned to collections of two-level atoms. We
employ the main concept of Statistical Mechanics that asserts that states of isolated systems that correspond to the
same total energy (degenerate states) are equally likely to occur. 

Let $n$ denote the number of balls of weight 1 and $N-n$ the number of balls of weight 0. The energy is $E=n\epsilon$.
The number of possible configurations (degeneracy) is given by the formula (see, e.g, \cite{schroeder,arnaud2}) 
\begin{equation}
\label{mathcalW}
\mathcal{W}(n) = \frac{N!}{n!(N-n)!} .
\end{equation}
For $N=3$ and $n=1$ for example, the possible configurations are (1,0,0), (0,1,0), and (0,0,1), where we have listed in
the parentheses the weights of the first, second and third balls. The degeneracy is therefore $\mathcal{W}(1)=3$ in
agreement with (\ref{mathcalW}).  The entropy is defined as: $S(E)=\ln(\mathcal{W}(n))$ if the Boltzmann constant is set
equal to unity, and the inverse temperature as: $\beta\equiv dS(E)/dE= [d\ln(\mathcal{W}(n))/dn]/\epsilon$. The
derivative cannot be evaluated exactly since $n$ is an integer.  The best one can do is to apply the Stirling
approximation of factorials: $d\ln(n!)/dn\approx \ln(n!)-\ln((n-1)!)=\ln(n)$. We obtain from \eqref{mathcalW}
\begin{equation}
\label{exp}
\beta = \frac{\ln(\frac{N}{n}-1)}{\epsilon} \iff
\exp(-\beta\epsilon)=\frac{n}{N-n} \iff
f \equiv \frac{n}{N}=\frac{1}{\exp(\beta\epsilon)+1}.
\end{equation}
The second relation says that the Boltzmann factor $\exp(-\beta\epsilon)$ is the ratio of the number $n$ of balls of
weight 1 and the number $N-n$ of balls of weight 0. The reservoir temperature is positive when $n<N/2$ and negative when
$n>N/2$.

\bibliography{Simple}

\begin{thebibliography}{10}
\bibitem[{not({\natexlab{a}})}]{note1}
\bibinfo{note}{The zero of energy is often taken half-way between the system
  lower and upper levels. Slight changes in the formulas would result from this
  alternative convention.}
\bibitem[{not({\natexlab{b}})}]{note2}
\bibinfo{note}{The operation of classical heat engines with negative bath
  temperatures has been treated by Landsberg \cite{landsberg}.}
\end{thebibliography}

\end{document}